\documentclass[prl, twocolumn, showpacs, amsmath]{revtex4}
\usepackage{amsmath}
\usepackage{amsthm}
\usepackage{amsfonts}
\usepackage{amssymb}
\usepackage{graphicx}
\theoremstyle{definition} \newtheorem{defi}{Definition}
\theoremstyle{plain} \newtheorem*{theo}{Theorem}
\theoremstyle{remark} \newtheorem*{side}{Proof}

\begin{document}


\title{On the Clausius formulation of the second law in stationary chemical networks through the theorems of the alternative.}


\author{Daniele De Martino}
\affiliation{Dipartimento di Fisica, Sapienza Universit\`a di Roma,p.le A. Moro 2, 00185 Roma (Italy)}



\begin{abstract}
In this article the Gordan theorem is applied 
to the thermodynamics of a chemical reaction network at steady state.
From a theoretical viewpoint it is equivalent to the Clausius formulation 
of the second law for the out of equilibrium steady states of chemical networks, 
i.e. it states that the exclusion (presence) of closed reactions loops
makes possible (impossible) the definition of a thermodynamic potential and vice versa. 
On the computational side, it reveals that calculating reactions free energy  
and searching infeasible loops in flux states are dual problems 
whose solutions are alternatively inconsistent.
The relevance of this result for applications is discussed with an example
in the field  of constraints-based modeling of cellular metabolism where 
it leads to efficient and scalable methods 
to afford the energy balance analysis. 
\end{abstract}


\maketitle

\section*{Introduction}
The non-equilibrium thermodynamics of chemical reaction networks
has been a subject of great interest and research efforts in recent years,
from the pioneering works on network thermodynamics\cite{perelsonet} to the  
recent study of their statistical grounds through fluctuation theorems\cite{gaspard,Beard:2007uq}.
The quest for general results or underlying variational principles\cite{dualityprl} 
is open, as it is for out of equilibrium thermodynamics in general\cite{kondeprigo}.
Regarding its applications, a chemical reaction network at steady state
under the hypothesis of local equilibrium and well-mixing 
is the simplest (and standard) model of cell metabolism.
This model has been applied to large (genome-) scale systems only in recent times,
where it was successful in describing and predicting, even at a quantitative level, 
cellular metabolic processes\cite{Varma:1994kc}.
In general, the stochiometry of the network defines linear mass-balanced equations 
for the reaction fluxes and thus a feasible solution space for the steady states.    
In particular, putting forward a functional hypothesis, one solution can be selected    
by maximizing a linear function  that describes e.g. the biomass or ATP production. 
In this way is possible to use powerful methods and tools from linear programming,
a framework that goes under the name of flux balance analysis (FBA)\cite{Orth:2010if}.
More recently, the question of the implementation of the thermodynamic constraints 
has been posed, a framework has been called energy balance analysis (EBA)\cite{Beard:2002vn}.  
This problem is connected on one hand to the calculation of reactions free energy, 
metabolites chemical potential and concentration\cite{Kummel:2006kl,Henry:2007xz},
on the other to the removal of infeasible loops from flux configurations\cite{Price:2002mw}. 
All through the literature there is the intuition of the infeasibility 
of closed reaction loops, since it is easy to show that their existence renders 
impossible to define a thermodynamic potential. 
So, their exclusion is necessary, but is it \emph{sufficient} 
to guarantee the thermodynamical consistency?
In this article a positive aswer to this question will be given, 
showing that it comes from the application of the Gordan theorem, 
the oldest theorem of the alternatives and a key result in optimization\cite{linearpro}.
The application of the Gordan theorem in this context states shortly that, 
under the hypothesis of local equilibrium for well-mixed systems at constant pressure and temperature,
a consistent thermodynamic potential can be defined \emph{if and only if} there are no closed loops.
This is substantially equivalent to the Clausius formulation of the second law in networks where closed
reaction loops play the role of \emph{perpetuum mobile of the second kind} and  
its demonstration leads to address more efficiently the energy balance analysis of metabolic networks.
Infact, more in details, the theorem reveals that finding infeasible reactions loops 
and/or calculating reactions free energies are dual problems whose solutions are alternatively inconsistent.
For instance, current methods apt to find and eliminate infeasible loops suffers of scalability issues\cite{Schellenberger:2011uq}. 
Then, this difficulty can be circumvented, by dealing with the easier dual problem of calculating reactions free energy, 
that consists in solving a system of linear inequalities.
In the following we will give an elementary demonstration of this theorem and discuss 
its applications with an example in the field of constraints-based modeling of cellular metabolism 
where it gives efficient methods to afford the energy balance analysis.

\section{Results}
Suppose the stochiometry of a chemical reaction network is given 
in terms of a matrix $\mathbb{S}\equiv (S_i^\mu)$, where $S_i^\mu$ stands for the stochiometric coefficient 
of the chemical species $\mu = 1 \dots M$ in the reaction $i=1 \dots N$. 
A flux vector ${\bf v} \equiv (v_i)$ is by definition\cite{Beard:2008fk} \emph{thermodynamically feasible} provided the existence 
of a free energy vector ${\bf \Delta G}\equiv (\Delta G_i)$ such that
\begin{defi}[Thermodynamic feasibility]
\begin{eqnarray}
\label{def}
v_i \Delta G_i < 0 \qquad \forall i   \nonumber \\
{\bf r}\cdot {\bf \Delta G}=0  \qquad \forall {\bf r}\in Ker(\mathbb{S}).
\end{eqnarray}
\end{defi}  
Writing the free energy change in terms of the chemical potentials ${\bf g}\equiv (g_\mu)$
\begin{equation}
\Delta G_i = \sum_\mu S_i^\mu g_\mu,
\end{equation}
and defining for sake of semplicity in notations $\xi_i^\mu  = -sign(v_i) S_i^\mu $, 
we have a system of linear inequalities equivalent to (\ref{def}) 
\begin{equation}
\label{sys1}
\sum_\mu \xi_i^\mu g_\mu > 0.
\end{equation}
Now, the existence of solutions of the system (\ref{sys1}) is ruled by the Gordan theorem: 
the system (\ref{sys1}) has solution  
\emph{if and only if} the dual system
\begin{defi}[Infeasible loops]   
\begin{eqnarray}
\label{sys2}
\sum_i \xi_i^\mu k_i = 0 \qquad \forall \mu  \nonumber \\
k_i\geq 0, \quad {\bf k}\neq {\bf 0}
\end{eqnarray}
\end{defi}
has no solution, where the solutions of (\ref{sys2}) define closed reaction loops.
In other terms, 
\begin{theo}[Gordan theorem]
One and only one of the systems (\ref{sys1}) and (\ref{sys2}) have solution.
\end{theo}
\begin{side}
(\ref{sys2}) has solution $\Rightarrow$ (\ref{sys1}) has no solution.
\end{side}
Suppose to have a solution ${\bf k^*}$ of (\ref{sys2}). Take any vector ${\bf g}$, multiply it component by component
with the equations of the system (\ref{sys2}) and sum over them. Exchanging the indeces $\mu$ and $i$ 
in the sums and given the positivity of ${\bf k^*}$, 
it is straightforward to conclude that no vector ${ \bf g}$ can satisfy the system (\ref{sys1}). 

\begin{side}
(\ref{sys1}) has no solution $\Rightarrow$ (\ref{sys2}) has solution.
\end{side}

The demonstration is given by induction in the number $M$ of unknows.

The statment is true for $M=1$. Infact, for one unkonwn, the system (\ref{sys1}) is inconsistent  
if and only if there is at least one couple  $i$, $j$ such that $\xi_i^1 \xi_j^1 = -1/c < 0$, 
and in this case $k_i=1$, $k_j=c$ and $k_l=0 \quad \forall l\neq i,j$ is a solution of (\ref{sys2}).    
 
Let's consider a system of the type (\ref{sys1}) with $M$ unknows and suppose it is inconsistent.
We will prove that the dual system has solutions supposing the theorem true for systems with $M-1$ unknows.     
We have $\forall i \quad \sum_{\mu=1}^{M-1} \xi_i^\mu g_\mu > -\xi_i^M g_M$, 
 if $\xi_i^M \neq 0$, we can define $\tilde{\xi}_i^\mu = - \xi_i^\mu/\xi_i^M$, and we have:
\begin{eqnarray}
\sum_{\mu}^{M-1}\tilde{\xi}_i^\mu g_\mu = P_i > g_M  \qquad  \forall i: \quad \xi_i^M<0 \nonumber \\
\sum_{\mu}^{M-1}\tilde{\xi}_j^\mu g_\mu = Q_j < g_M  \qquad  \forall j: \quad \xi_j^M>0 \nonumber \\
\sum_{\mu}^{M-1}\xi_l^\mu g_\mu = R_l > 0   \qquad \forall l: \quad \xi_l^M=0.
\end{eqnarray}  
Writing the system in this form, we can pass to the following system in $M-1$ unknows:
\begin{eqnarray}
\label{sysprof}
P_i> Q_j \qquad \forall i,j: \quad \xi_i^M<0 \quad \xi_j^M>0 \nonumber \\
R_l > 0   \qquad \forall l: \quad \xi_l^M=0.
\end{eqnarray}
Now, also this system is inconsistent. \\ Suppose infact there is a solution ${\bf g^*} = ( g_\mu^* )$, $\mu=1 \dots M-1$.\\ 
We could add to it any $g_M^*$ such that   $max_j Q_j({\bf g^*}) < g_M^* <min_i P_i({\bf g^*})$ 
and we will have a solution for the original system as well, against the hypothesis.     

By induction hypothesis, the theorem is true for systems with $M-1$ unknows. 
Then, referring to (\ref{sysprof}), there are $\tilde{k}_{ij}\geq 0$, $k_l \geq 0$ with at least one positive,  
such that $\sum_{ij} \tilde{k}_{ij} (\tilde{\xi}_i^\mu - \tilde{\xi}_j^\mu) + \sum_l k_l \xi_l^\mu = 0 \quad \forall \mu$.
From this we have finally a solution for the system (\ref{sys2}):
\begin{eqnarray}
k_i = - \sum_j \tilde{k}_{ij}/\xi_i^M  \qquad  \forall i: \quad \xi_i^M<0 \nonumber \\
k_j = \sum_i \tilde{k}_{ij}/\xi_j^M \qquad   \forall i: \quad \xi_j^M>0  \nonumber \\
k_l                 \qquad    \forall l: \quad \xi_l^M=0,
\end{eqnarray}
and the theorem is proven. 
For sake of simplicity the case in which some of the fluxes are null is neglected, 
but it follows along the same lines as an application of the Motzkin theorem\cite{linearpro}.

Apart from its inherent theoretical interest, this result can be used to perform 
efficiently the energy balance analysis of metabolic networks, 
because it suggests to deal with the dual system (\ref{sys1})
if we want to find and correct flux configurations from infeasible loops 
(calculating and/or correcting a consistent chemical potential vector as a by-product).

\begin{figure}[h!]
\begin{center}
\includegraphics*[width=.4\textwidth,angle=0]{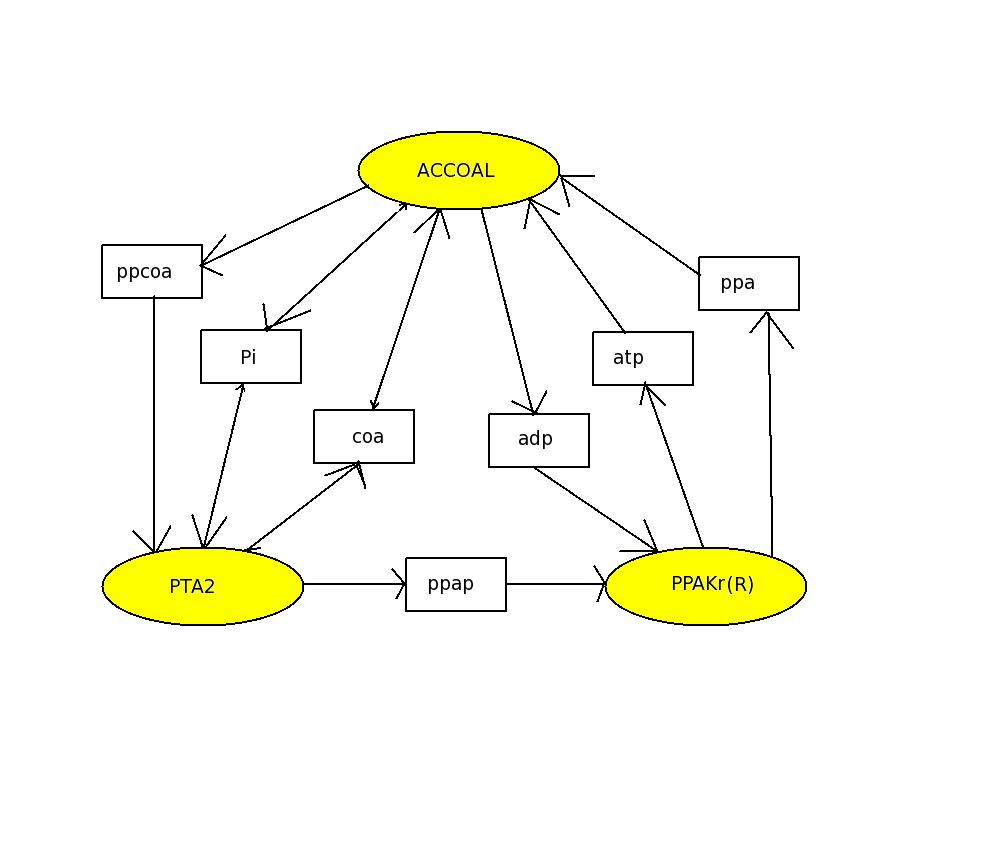}
\caption{An infeasible loop that can be present
 in the metabolic network model iJ904 if the putative reversible reaction PPAKr is considered to work in the backward direction.}
\label{loop}
\end{center}
\end{figure}

An algorithm was provided in\cite{noiplos} complementing 
standard relaxation method\cite{linearpro} (or MinOver scheme\cite{minover0}) with an exaustive search,
in the spirit of recent constructive demonstrations of the Farkas Lemma\cite{FMdemo}, 
the most famous extension of the Gordan theorem.
The  algorithm works by correcting step by step the least unsatisfied constraint, 
$\alpha$ being constant (Minover) or proportional to the amount by which the constraint 
is violated (Relaxation), i.e. a series is defined:
\begin{eqnarray}
i_l = min_i \sum_\mu \xi_i^\mu g_\mu(l) \\
g_\mu(l+1) = g_\mu(l) +\alpha \xi_{il}^\mu,
\end{eqnarray}  
till a solution is found, if any, in polynomial time\cite{noiplos}.
If the algorithm doesn't converge, by the theorem there should be infeasible loops.
Then, looking at the series of distinct least unsatisfied constraints, 
that unless of very bad cases is not large, it is possible to find them.
The Minover scheme alone was already applied to the dual part of the stochiometric matrix
of metabolic networks in order to calculate the production profile\cite{Martelli:2009jl}.
In \cite{noiplos} this algorithm was applied to calculate   
the list of infeasible loops in the recent model of E.coli metabolism iAF1260,
here, as a further example, it is reported in Table \ref{loops} a list of the possible unfeasible loops  
founded with this method in the model of E.Coli metabolic network iJR904\cite{Reed:2003ws}. 
This network is composed of $931$ reactions ($245$ of which are reversible) among $761$ metabolites. 
One of this loop is depicted in fig.\ref{loop}, from which it is clear that e.g. the putative reversible reaction
PPAKr sohuld operate in the backward direction in order to guarantee thermodynamic feasibility.
The thermodynamic feasibility of flux configurations in which the presence of these loops is removed from the beginning 
was tested over $10^6$ random instances.  
The small number of such loops should be ascribed to the detailed prior thermodynamic information
on reaction reversibility usually provided with the network models, 
whose assessement comes from a careful estimate of the chemical potentials\cite{Fleming:2009qa}. 

\begin{table}
\begin{center}
\begin{tabular}{| c | c | c | }  
\hline 
Cycle ID & Lenght & Formula  \\ 
\hline 
1&3  &  ACCOAL  $+$     PTA2  $-$     PPAKr(R)   \\
2&3  &  GLUABUTt7(R)  $-$     GLUt2r(R)  $+$     ABUTt2   \\
3&3  &  ACCOAL $-$   SUCOAS(R)  $+$      PPCSCT   \\
4&3  &  VALTA $-$ ALATA\_L(R)  $+$     VALTA    \\
5&3  &  ADK3(R)  $-$ ADK1(R)  $+$     NDPK1(R)    \\
6&3  &  ADK1(R) $-$ ADK3(R) $-$     NDPK1(R)    \\
7&3  &  NAt3\_1(R)  $-$ THRt2(R)   $+$     THRt4  \\
8&3  &  NAt3\_1(R)  $-$     PROt2r(R)  $-$   PROt4   \\
9&3  &  NAt3\_1(R)  $-$     GLUt2r(R)  $+$     GLUt4  \\
10&3  &  NAt3\_1(R)  $-$     SERt2r(R)  $+$     SERt4  \\
\hline
\end{tabular}
\caption{The 10 unfeasible cycles identified from thermodynamic feasibility analysis of $10^6$ different randomly generated flux configurations 
for the {\it Escherichia coli} metabolic reaction network iJR904. 
Plus (resp. minus) signs indicate that the reaction participates in the cycle in its forward (resp. backward) direction. 
(R) indicates that the corresponding reaction is putatively reversible according to \cite{Reed:2003ws}.  
The directions of the putatively reversible fluxes PPAKr and SOCOAS turn out to be in fact constrained.}
\label{loops}
\end{center}
\end{table}

\section*{Conclusions} 
In this article was given a rigourous proof that reaction fluxes in a chemical reaction network at steady state 
are thermodynamically feasible if and only if there are no closed reaction loops. 
It was showed that this is not a trivial statment but an application of the Gordan theorem.
Infact, it states in detail that calculating free energies and/or chemical potentials 
and assessing the thermodynamic feasibility of flux configurations 
through the search for infeasible loops are dual problems whose solutions are mutually inconsistent.
This in turn provides efficient ways to perform the energy balance analysis of metabolic networks and a genome-scale example was given. 
From a theoretical viewpoint, the result could be extended to a more general set of  processes 
to define the underlying thermodynamic variational principle itself:
since there are no loops, a potential can be defined whose changes have a definite sign.
From this point of view the theorem extends the Clausius formulation of the second law in a network context, 
that could be an interesting result for the emerging field of network science\cite{barbabasi} as well.   
Further, it would be interesting to investigate the statistical basis of this theorem.
Regarding applications, apart from thermodynamic feasibility, another important problem related to flux analysis
in constraint based modeling concerns the link between the network structure and its productive capabilities\cite{Kyoto10}.
In a recently developed framework inspired by economic systems modeling
\cite{Martino:2004il}, the fluxes are calculated from the minimal constraint 
that for each chemical species at stationarity the overall consumption cannot exceed the total supply (Von Neumann constraints),
and the aim is to calculate producibility profiles and to infer the biomass coefficients directly from the stochiometry\cite{Martelli:2009jl}.
The application of the the Farkas lemma in this context, i.e. to the dual part of the stochiometric matrix, 
has been partially shown in \cite{imielinski} in order to study the growth media for E.Coli, 
revealing the connection between producibility and conserved metabolic pools.
However, similarly to this work, these theorems of the alternative could provide also interesting insights for the calculation itself
of the producibility profile toghether with the conserved metabolic pools, an aspect that is leaved for future investigations.  

\section*{Acknowledgments}

This work is supported by the DREAM Seed Project of the Italian Institute of Technology (IIT), 
by the IIT Computational Platform and by the joint IIT/Sapienza Lab ``Nanomedicine''. 
The author thanks A. De Martino, M. Figliuzzi and E. Marinari for useful suggestions, fruitful discussions and encouragement.

\bibliography{thermobib}

\end{document}